 \documentclass{agujournal}

\journalname{JGR-Space Physics}
\usepackage{hyperref}

\begin{document}

%
%

\title{Pulsations in the Earth's Lower Ionosphere Synchronized with Solar Flare Emission}
%
%




\authors{Laura A. Hayes\affil{1,2,3}, Peter T. Gallagher\affil{1}, Joseph McCauley\affil{1},\\ Brian R. Dennis\affil{2}, Jack Ireland\affil{2,3}, Andrew Inglis\affil{2,4}}


\affiliation{1}{ School of Physics, Trinity College Dublin, Dublin 2, Ireland.}
\affiliation{2}{Solar Physics Laboratory, Heliophysics Science Division, NASA Goddard Space Flight Center, Greenbelt, MD 20771, USA.}
\affiliation{3}{ADNET Systems, Inc., USA}
\affiliation{4}{Physics Department, The Catholic University of America, Washington, DC 20664, USA }



\correspondingauthor{Peter T. Gallagher}{peter.gallagher@tcd.ie}



\begin{keypoints}
\item Pulsations detected in ionospheric D-region are synchronized with flare X-ray pulsations.
\item D-region electron density varies by up to an order of magnitude over 20 minutes during the pulsations.
\item Effective ionospheric recombination coefficient can be calculated as a function of varying X-ray flux.
\end{keypoints}

Paper Accepted to JGR Space Physics. DOI: 10.1002/2017JA024647
%
%


\begin{abstract}
Solar flare emission at X-ray and extreme ultraviolet (EUV) energies can cause substantial enhancements in the electron density in the Earth's lower ionosphere. It has now become clear that flares exhibit quasi-periodic pulsations with timescales of minutes at X-ray energies, but to date, it has not been known if the ionosphere is sensitive to this variability. Here, using a combination of Very Low Frequency (24 kHz) measurement together with space-based X-ray and EUV observations, we report pulsations of the ionospheric D-region, which are synchronized with a set of pulsating flare loops. Modeling of the ionosphere show that the D-region electron density varies by up to an order of magnitude over the timescale of the pulsations ($\sim$ 20 mins). Our results reveal that the Earth's ionosphere is more sensitive to small-scale changes in solar soft X-ray flux than previously thought, and implies that planetary ionospheres are closely coupled to small-scale changes in solar/stellar activity.
\end{abstract}

%
%

%


%
%
%
%

\section{Introduction}
All planetary atmospheres respond to solar flare activity \citep{witasse}. On Earth, sudden increases in extreme ultraviolet (EUV) and X-ray radiation during a solar flare affect the entire dayside ionosphere \citep{tsurutani2009brief}. During quiet Sun conditions, the D-region ($\sim$60-90~km in altitude) is maintained by Lyman-$\alpha$ (1216~\AA) acting on the minor constituent nitric oxide, with the X-ray flux being too small to be a contributor. However, when a solar flare occurs, X-ray photons of wavelengths <10~\AA\ can penetrate down to the D-region resulting in a dramatic increase in ionization of all neutral constituents (including nitrogen and oxygen) in this lowest lying region of the Earth's ionosphere \citep{mitra74, whitten1965physics}. This markedly increases the electron density of the D-region. Observations of the propagation characteristics of very low frequency (VLF; 3-30 kHz) radio waves  provide a tool to remotely investigate the behavior of this ionospheric layer in response to ionizing disturbances \citep{mitra74,thomson2005large}. VLF waves propagate in the waveguide bounded below by the Earth's surface and above by the lower ionosphere. Enhancements in ionization of the D-region result in amplitude and phase variations of the received VLF signal. Theoretical models \citep{wait1964, budden1988propagation, ferguson1998computer} can then be used in conjunction with VLF measurement to estimate the variation of electron density with height and time  through the D-region, which are not readily measurable by  other means. Indeed, VLF techniques have also been used in the detection of transient ionospheric disturbances at D-region altitudes from phenomena such as gamma-ray bursts \citep{inan1999}, solar eclipses \citep{kumar2016changes}, space weather \citep{kumar2015response}, meteor showers \citep{kaufmann1989effects}, and from disturbances such as lightening strikes \citep{inan2010_lightening, nat_ion}, and earthquakes \citep{earthquake}.

It has now become clear that X-ray emission generated in solar and stellar flares show pronounced pulsations and oscillatory behavior, with periods ranging from seconds to several minutes \citep{nak2009}.  The underlying mechanism responsible for producing  these `quasi-periodic pulsations' (QPP) in flaring emission remain debated. One possibility is that the pulsations are a direct result of multiple bursts of energy release from a regime of periodic magnetic reconnection. This may explain the `bursty', shorter period (seconds) pulsations observed during the impulsive phase of solar flares but fail to account for the longer period QPP.  Longer period pulsations are generally assumed to be a manifestation of magnetohydrodynamic (MHD) wave processes excited in the corona and/or flare sites. Pulsations, with periods of minutes to tens of minutes have previously been reported in the X-ray emission from flaring events \citep[e.g.][] {vsvestka1982unusual,  harrison1987solar, li}. It has been suggested that slow-mode oscillations of large scale loops could drive the observed pulsations \citep{vsvestka1994slow}. Alternatively \citep{foullon2005} interpreted X-ray QPP with periods of 8-12 minutes as periodic modulation of reconnection by an external MHD oscillation in a nearby loop. Similar timescale pulsations have also been studied in EUV, so called SUMER oscillations, and interpreted as standing longitudinal slow mode waves in hot coronal loops \cite{wang}.

It has recently been noted that these long-period time-scales are a common feature of soft X-ray emission observed in the 1-8~\AA\ channel of the Geostationary Operational Environment Satellite (GOES) \citep{tan2016very}. The question now arises as to whether these oscillatory signatures of soft X-ray pulsations can appear in terrestrial observables of our ionosphere.  To the best of our knowledge there has been no reports in the literature of solar flare quasi-periodic pulsations and their response in the lower ionosphere. Here, using EUV imaging from the Atmospheric Imaging Assembly (AIA) aboard the Solar Dynamics Observatory (SDO) together with the GOES X-ray sensor and VLF monitoring of the lower ionosphere we reveal previously unseen characteristics of the relationship between dynamic oscillatory signatures in both the solar atmosphere and the Earth's ionosphere.

\section{Flaring Pulsations and Ionospheric Response}

On 24 July 2016, an active region located on the western limb of the Sun began to flare (NOAA active region 12567; N05W91; see Figure 1a--c). Over the course of four hours, from 11:00--15:00 UT, a series of X-ray pulsations of GOES class B9.2--C6.8 were observed in the  1--8 \AA\ band of the GOES X-ray sensor. The flaring loops, imaged using EUV observations in the hot 131~\AA\ AIA passband  (peak response $\sim$10~MK) \citep{lemen}, similarly reveal quasi-periodic brightness variations (see Supplementary Movie 1). The evolution of the X-ray emission together with the associated AIA 131~\AA\ lightcurve integrated over the active region is shown in Figure 1d. The time profiles exhibit quasi-periodic pulsations with a progression of 9 large amplitude peaks of growing intensity. Periodogram analysis of the flux profiles finds a characteristic timescale of $\sim$20 minutes between peaks of the pulsations.

\begin{figure}[h!]
\centering
\includegraphics[width =0.90\textwidth]{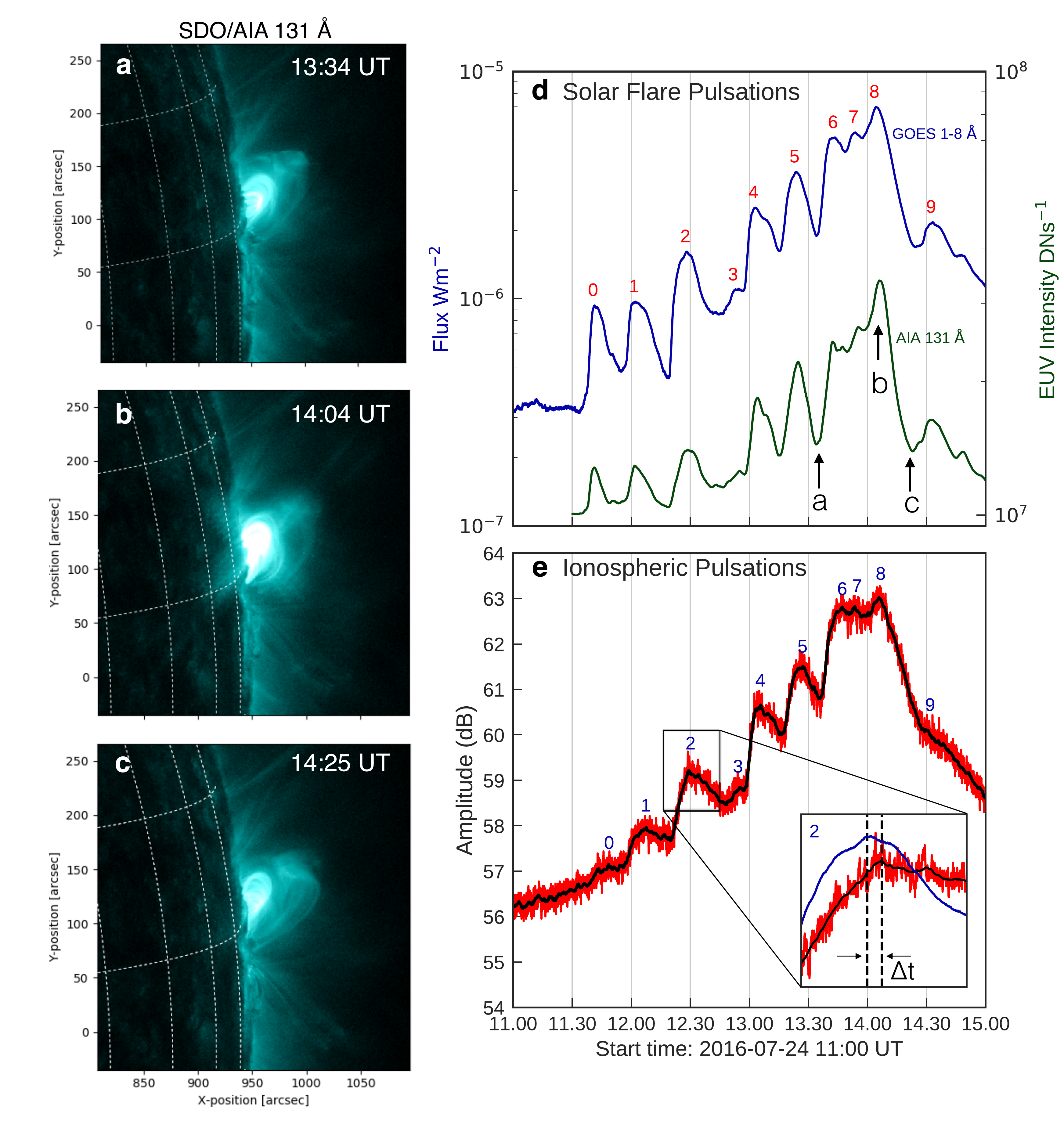}
\caption{Synchronized pulsations from coronal flaring loops observed in both X-ray and EUV emission and the response of the D-region of the ionosphere. EUV images from the 131~\AA\ channel of SDO/AIA is shown in a, b and c during the three intervals marked by the arrows in d. Quasi-periodic pulsations are evident in both the X-ray and EUV emission in (d), while the corresponding D-region response observed using VLF at 24~kHz is shown in (e). The subplot in (e) is a zoom in of the pulsation numbered 2 to highlight the time delay ($\Delta$t $\sim$ 90s) between the X-ray peak and the VLF response.}
\label{EUV_flare}
\end{figure}

To examine the lower ionosphere response to  X-ray QPP,  VLF radio signals at 24 kHz emitted by the communications transmitter in Maine, US  (station id: NAA; 44.6$^{\circ}$N 67.2$^{\circ}$W) were monitored at the Rosse Solar--Terrestrial Observatory in Birr, Ireland (53.1$^{\circ}$N, 7.9$^{\circ}$W) using Stanford University Sudden Ionospheric Disturbance (SID) monitor \citep{scherrer2008distributing}. This  propagation path had a great circle distance of $\sim$5,320 km across the Atlantic Ocean, which provided a continuous sunlit path to remotely measure the response of the ionosphere as the Sun traversed the ocean (Figure 2). The influence of the X-ray pulsations on the lower ionosphere can clearly be identified in the received VLF amplitude (dB) (Figure 1e). The enhancement in amplitude of the VLF signal results from increased electron density in the D-region, which lowers and sharpens the upper mirror point of the Earth-ionosphere waveguide, allowing the VLF signal to reflect at a sharper boundary with less attenuation \citep{thomson2001solar, grubor05}.

Notably, the amplitude of the received VLF signal exhibits pulsating signatures that systematically track the pulsations in X-ray emission. A Pearson correlation coefficient of 0.92 was found between the GOES 1-8 \AA\ time profile and the VLF response. Taking into account the diurnal solar zenith angle variation due to Lyman $\alpha$ emission from the Sun, we also compute the differential VLF response and compare with the X-ray pulsations. This is done by subtracting the mean of solar quiet days close to the event. The correlation coefficient of the differential VLF and the X-ray flare is then found to be 0.94. The fact that soft X-ray quasi-periodic pulsations produce synchronized pulsations in the D-region electron density (as monitored by the VLF response) indicate a close coupling of the solar-terrestrial relationship as X-ray QPP act as an external quasi-periodic driver to the D-region electron density. Each peak in X-ray and VLF response is numbered 0-9 for comparison. The signal response of the VLF to the X-ray flux is more significant with each pulsation. This is due to the increasing X-ray intensity with each pulse. The electron production rate ($q$) is directly proportional to the flux ($F_s$~ Wm$^{-2}$) of ionizing radiation \citep{ratcliffe, budden1988propagation}, hence we see larger amplitude responses for greater X-ray flux. 

A common feature of the observed pulsations is the time delay ($\Delta t$) between the peak of the VLF amplitude and the peak of X-ray flux. This time delay is highlighted in the sub panel in Figure 1e, where a zoomed plot of both the X-ray lightcurve and VLF response is shown for peak 2. Cross-correlation analysis between profiles finds a delay of $\sim$90 seconds. The time delay present here is a characteristic feature of the response of the ionosphere to ionizing radiation. Previous authors have noted it as the `relation time' \citep{mitra74} or `sluggishness' \citep{appleton} of the ionosphere in response to ionizing flux.  In the lower ionosphere, the electron production rate ($q = F_{s} \sigma N_{e}$~m$^{-3}$ s$^{-1}$) is dominated by photoionization, while electron losses ($L = \alpha_{eff} N_{e}^2$~m$^{-3}$ s$^{-1}$) result from recombination \citep{mitra74, zigman2007, nina2012altitude}.  The physical effect of the electron loss process is to delay the response of the changes in electron density $N_e$~(m$^{-3}$) to changes in $F_s$.  Here, $\alpha_{eff}$~(m$^{3}$ s$^{-1}$) is the effective recombination coefficient. Hence, there is a time delay between the ionizing X-ray peak and the VLF response and it signifies the time taken for the D-region photoionization-recombination processes to recover balance after increased irradiance \citep{zigman2007, appleton, basak2013effective}. 

\begin{figure}[h!]
\centering
\includegraphics[width =0.51\textwidth]{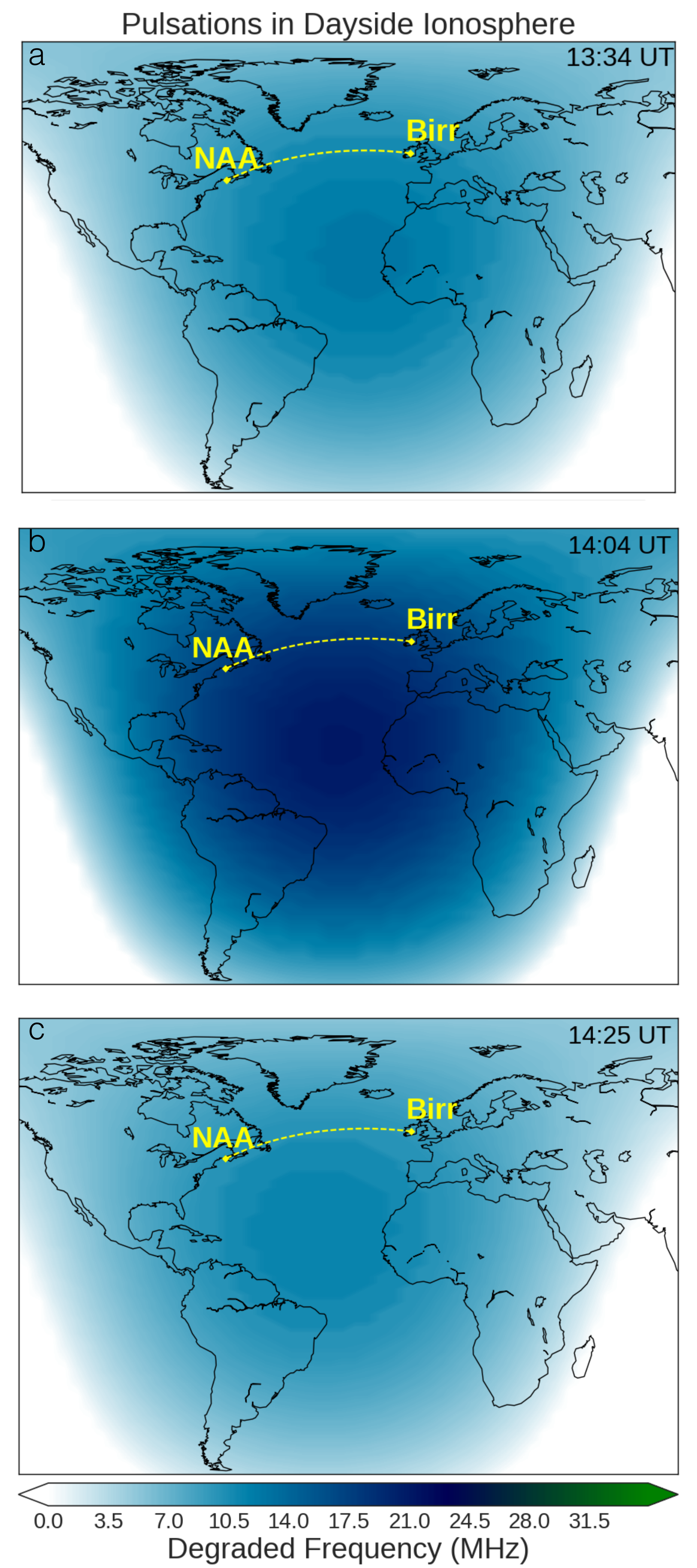}
\label{drap}
\caption{High frequency radio wave absorption as a result of increased ionization in the D-region of the ionosphere. The D-RAP model, driven by GOES X-ray and proton fluxes, show that the day-side of the ionosphere over the Atlantic Ocean suffered increased ionization during the solar X-ray pulsations. The three panels a, b and c correspond to the times shown in Figure 1. a, b, and c respectively. See also Supplementary Movie 2.}
\end{figure}

\section{D-Region Absorption Model}
The D-region Absorption Model (D-RAP) from the NOAA Space Weather Prediction Center \citep{drap_doc} provided additional insight into the effects of the X-ray pulsations on the terrestrial ionosphere.  Although the excess ionization at the D-region improves reflectivity at low frequency radio waves, it has deleterious impacts on higher frequency sub-ionospheric radio communications. High frequency (HF; 3--30 MHz) waves allow long-distance communication as they reflect in the upper ionosphere (peak of F2 region), passing through the D-region as they propagate. When a strong X-ray flare occurs, HF signals suffer attenuation due to absorption as it passes through the increased local electron density of the D-region. In extreme cases these signals can fade out as they are absorbed before and after they undergo reflection. D-RAP models this effect on HF wave attenuation attributed to X-ray enhancements in the D-region and was used here to illustrate the HF propagation affects over the path traveled by VLF (from Maine, USA to Birr, Ireland). The global D-RAP map is shown in Figure 2 at three intervals corresponding to times of different ionization enhancement (same time as Figure 1a, b, and c respectively). The map shows the great circle path from the NAA transmitter to our receiver at Birr, together with the highest frequency affected by absorption of 1dB. Throughout the flaring event, ionization perturbations take place in the D-region along this path illustrating that the NAA-Birr propagation is an appropriate path to use for VLF remote diagnostics in relation to this flaring event (See also Supplementary Movie 2).

\section{Modeling D-Region Electron Density}
To investigate the behavior of the electron population of the lower ionosphere during the QPP event, we use a full waveguide solution to model the propagation of the VLF signal during perturbed conditions. Our analysis is based on the Wait model \citep{wait1964}, which uses time dependent parameters of a reference height \textit{H'} (in km) and an electron density e-folding or `sharpness' $\beta$ (in km$^{-1}$) to describe the electron density height profile in the lower ionosphere. This two parameter system provides a model of a vertically stratified ionosphere with an electron density profile $N_{e}$ that increases exponentially with altitude, $h$: 

\begin{equation} 
\label{elec_eq}
N_e(h, H', \beta) = 1.43\times10^{13} e^{-0.15H'} e^{(\beta - \beta_0)(h-H')} \hspace{2mm}   \mathrm{m^{-3}}
\end{equation}

\noindent Here $\beta_0$ is 0.15~km$^{-1}$. The changes to this electron density height profile during  perturbed conditions were estimated using the Long-Waveguide Propagation Capability (LWPC) \citep{ferguson1998computer}  code. Specifying electron profile parameters (\textit{H'}, $\beta$), together with input path variables allows LWPC to simulate the VLF propagation based on full waveguide mode theory. It returns expected amplitude and phase at a specific receiving point. By varying the two independent parameters \textit{H'} and $\beta$, a range of VLF signal amplitudes were simulated, and the resulting agreement between modeled and observed signals provided the most likely height profile of the disturbed ionospheric.

\vspace{2mm}


\begin{figure}[h!]
\centering
\includegraphics[width =0.65\textwidth]{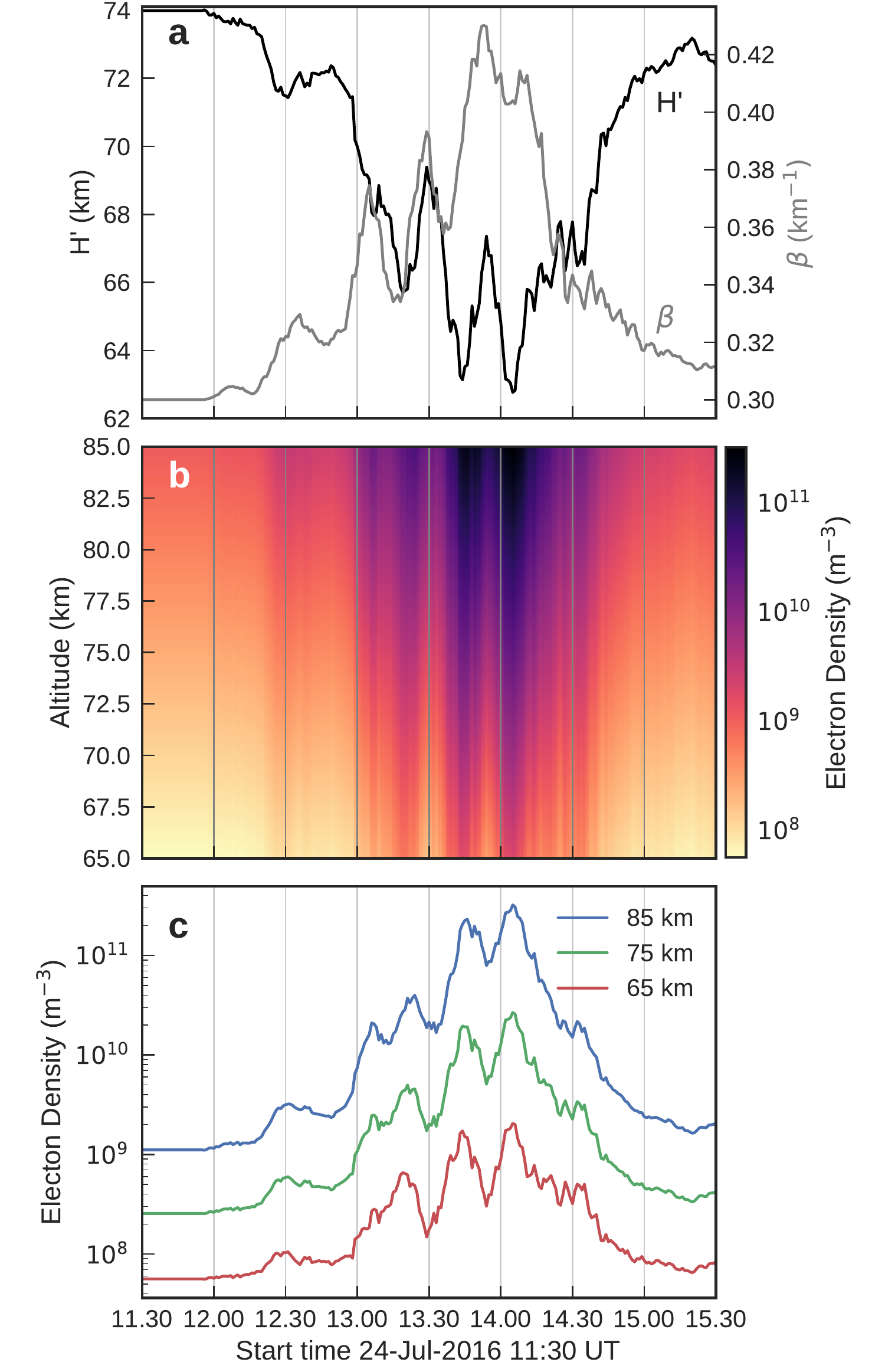}
\label{e_den}
\caption{Calculated Wait parameter values and variations in electron density in the lower ionosphere as a result of the pulsating X-ray flux. The set of $H'$ and $\beta$ values as a function of time computed from the LWPC simulation are shown in panel a. The electron density variations (in m$^{-3}$) is shown in panel b as a function of both altitude and time calculated from the electron density equation (1). Temporal electron density changes at altitudes of 65, 75 and 85~km are shown in panel c.}
\end{figure}

During quiet daytime conditions, the reference values for \textit{H'} and $\beta$ are 74$\pm$1 km and 0.31$\pm$0.01 km$^{-1}$, respectively. However, during the flare, the increased ionization lowers the effective height \textit{H'} and increases the sharpness parameter $\beta$. The temporal variations of these parameters are shown in Figure 3 a, with $H'$ and $\beta$ marked in black and gray respectively. As shown, the set of these parameters similarly follow the pulsating signatures. During the highest soft X-ray peak (at $\sim$ 14:05~UT), we find that \textit{H'} decreases from 74$\pm$1~km to 63$\pm$1~km, and $\beta$ increases from 0.31$\pm$0.01~km$^{-1}$ to 0.42$\pm$0.01~km$^{-1}$.  
The altitude and temporal dependence of the resulting modeled electron density calculated from equation (1) is displayed in Figure 3 b and c. During the flaring event, especially from peaks 2-8, the variation of the electron density is closely related to the soft X-ray ionizing flux, similarly displaying a pulsating signature. We can see that the influence of the flaring radiation is more pronounced at higher altitudes, and during times of larger X-ray intensity. For example, peaks 0 and 1 do show a temporal behavior in the ionosphere, more so at higher altitudes, however the magnitude of the variations is much smaller. At 74~km, the electron density increases from $\sim 2\times 10^8$ m$^{-3}$ during normal daytime conditions, to $\sim 1.6 \times 10^{10}$ m$^{-3}$ during the largest flare pulse. 

\vspace{2mm}

The electron loss processes can be quantified in terms of the effective recombination coefficient $\alpha_{eff}$. This coefficient can be related to the time delay ($\Delta t$) between the VLF response the X-ray flux. Using the electron continuity equation, $\alpha_{eff}$ can be calculated at a local maximum of the ionizing radiation (i.e., when $dq/dt = 0$) by the relation $\alpha_{eff} = 1 / (2 N_e \Delta t)$ \citep{zigman2007,mitra74,appleton}. Here $N_e$ is the electron density and $\Delta t$ is the time delay between VLF and X-ray peaks. Given the multiple pulsations in our example, we can calculate $\alpha_{eff}$ as a function of flux during the same flaring event. Using electron density values from Figure 3 at each peak for a selection of heights, together with a time delay of $\sim$90 seconds, we calculate the effective recombination coefficient at the times of soft X-ray peaks. The variation of $\alpha_{eff}$ as a function of X-ray flux, and altitude is shown in Figure 4.  The coefficient is calculated for each peak 0-9, as marked in the Figure. Unlike the electron density, the effective recombination coefficient has smaller values for larger X-ray flux, and at higher altitudes. The values in the range of 10$^{-10} - 10^{-13}$ m$^3$s$^{-1}$ are in agreement with other works \citep{zigman2007, gledhill}. The advantage of our analysis is that we can estimate multiple values for the effective recombination coefficient during the same flaring event. 
\begin{figure}[h]
\centering
\includegraphics[width =0.85\textwidth]{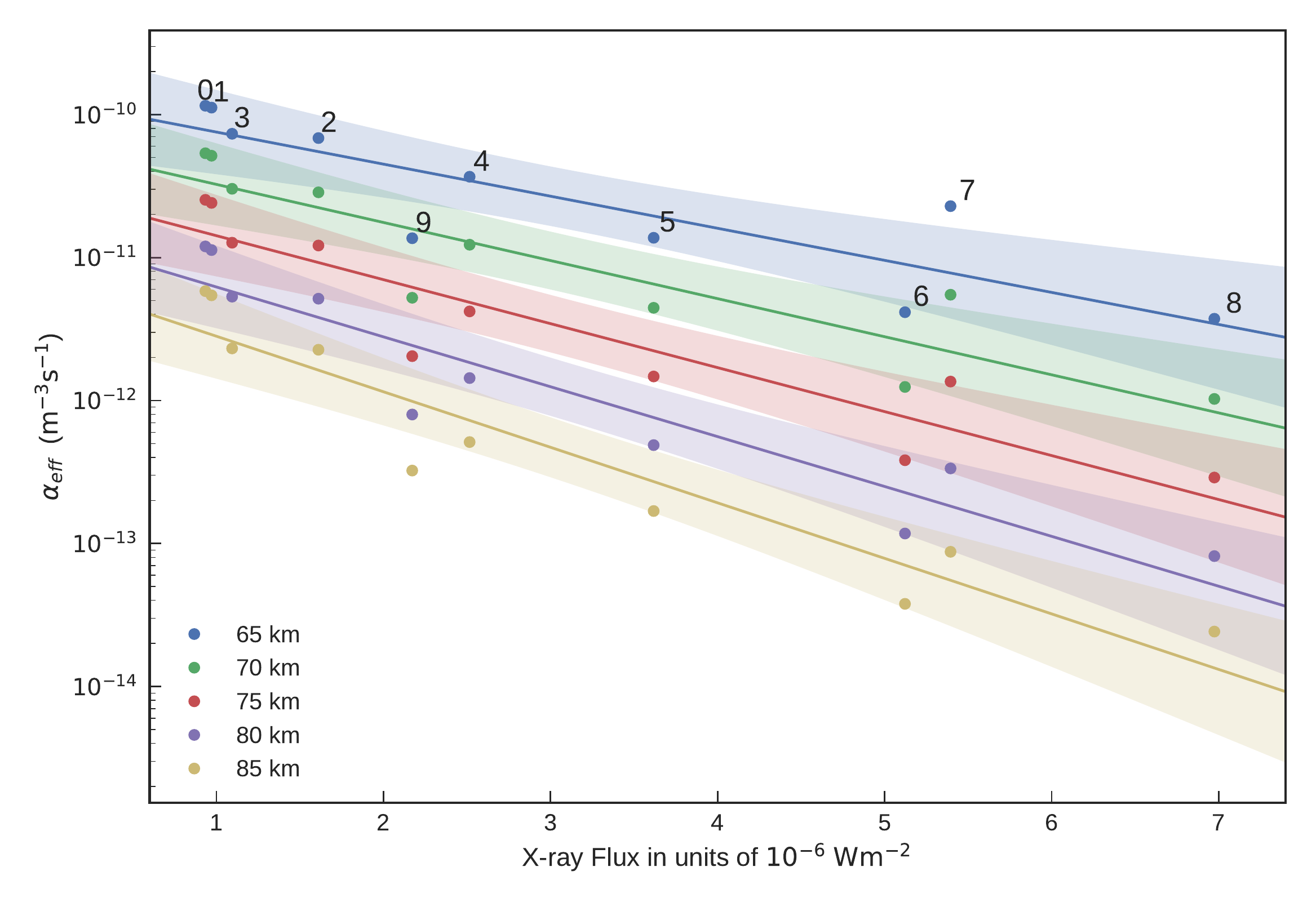}
\label{recombo}
\caption{Recombination coefficient $\alpha_{eff}$ as a function of X-ray flux and altitude. The coefficient is calculated at each X-ray peak time, numbered 0-9. The shaded areas are the 95\% confidence interval of the least-squares fit.}
\end{figure} 
\vspace{2mm}

\section{Conclusions}
The observations described here reveals that the lower ionosphere is sensitive to dynamic solar activity such as quasi-periodic pulsations in X-ray flaring emission. This is the first observation of synchronized pulsations, exposing a sensitive coupling of oscillatory signatures in solar flare soft X-ray emission and the Earth's ionized environment on  short time-scales. The detection of solar flare pulsations in the Earth's ionosphere has interesting implications for study of the geo-effectiveness of solar flares. The rapid variations of ionospheric plasma density (Figure 3) produced by the pulsating X-ray emission could result in further triggering of processes in the Earth's atmosphere. The density variation may constitute a periodic driver for dynamic atmospheric phenomena, such as acoustic gravity waves \citep{nina2013detection} . These effects would be greatly enhanced if a resonance occurs between the periodicity of the incoming flux and the natural frequencies of the ionosphere \citep{nakariakov2016magnetohydrodynamic}.

The comparative nature of solar flare X-rays and the ionospheric response (without pulsating signatures) have been investigated widely in the literature \citep[e.g.][]{ mitra74, thomson2005large, selv}. VLF provides a powerful tool to probe the lower ionosphere in response to ionizing agents, such as solar X-ray flares. The Earth's ionosphere essentially acts as a giant X-ray detector, responding characteristically to enhanced X-ray flux. Subsequently, the D-region has responded to the quasi-periodic pulsations present in the incident X-ray flux from this solar flare.  Global VLF networks, such as GFIDS \citep{gifds} will allow us to continuously monitor the response of the D-region to solar flare activity, and indeed will allow us to further explore the relationship of dynamic QPP in solar flares and their ionospheric response.

\vspace{2mm}
Like Earth, other planetary atmospheres are also significantly affected by solar flares. For example, Mendillo et al. 2006 \citep{mendillo06} showed that an enhancement of up to 200\% in electron density can occur in the Martian lower ionosphere in response to enhanced X-ray flux during a flare. Similar reports have also been shown in the studies of the effects of flares of the atmosphere of Venus \citep{kar}. The question now arises as to whether oscillatory signatures can be observed in other planetary atmospheres and how coupled the planetary atmospheres are to solar activity. Finally, given the recent evidence of QPP in stellar flare emission \citep{mitra2005first, cho2016comparison, pugh2016statistical}, this observation implies the presence of oscillatory responses in exoplanetary atmospheres.

\acknowledgments
This work has been supported by an Irish Research Council (IRC) Enterprise Partnership  Studentship between Trinity College Dublin (TCD), Ireland and Adnet System Inc., USA. We wish to thank the International Year of Heliophysics (IHY) 2007 program that provided TCD with the Stanford Sudden Ionospheric Disturbance (SID) monitor. The VLF data can be accessed on the Rosse Solar Terrestial Observatory website \url{www.rosseobservatory.ie}. We acknowledge the use of SDO/AIA data from \url{www.lmsal.com/get_aia_data}, GOES data from \url{www.swpc.noaa.gov/products/goes-x-ray-flux} and NOAA/D-RAP at  \url{www.ngdc.noaa.gov/stp/drap/data}. The data analysis in this report was carried out using  SunPy, an open-source and free community-developed solar data analysis package written in Python \citep{mumford2015sunpy}. We thank the anonymous referees for their valuable comments to improve the article.

%
%
%

\begin{thebibliography}{[10]}

\bibitem[{\textit{Akmaev et~al.}}(2010)]{drap_doc}
Akmaev, R. A., Newman, A., Codrescu, M., Schulz, C., Nerney, E. D-RAP Model Validation: I. Scientific Report, (2010).

\bibitem[{\textit{Appleton}}(1953)]{appleton}
Appleton, E. V. A note on the `sluggishness' of the ionosphere. \textit{Journal of Atmospheric and Terrestrial Physics} 3, 282-284 (1953).

\bibitem[{\textit{Basak and Chakrabarti}}(2013)]{basak2013effective}
Basak, T., Chakrabarti, S. K. Effective recombination coefficient and solar zenith angle effects on low-latitude D-region ionosphere evaluated from VLF signal amplitude and its time delay during X-ray solar flares. \textit{Astrophysics and Space Science} 348, 315-326 (2013). 

\bibitem[{\textit{Budden}}(1988)]{budden1988propagation}
Budden, K. G. The propagation of radio waves: the theory of radio waves of low power in the ionosphere and magnetosphere. \textit{Cambridge University Press}, 1988).

\bibitem[{\textit{Cho et~al.}}(2016)]{cho2016comparison}
Cho, I.-H., Cho, K.-S., Nakariakov, V. M., Kim, S., Kumar, P. Comparison of damped oscillations in solar and stellar X-ray flares. \textit{The Astrophysical Journal} 830, 110 (2016).



\bibitem[{\textit{Ferguson}}(1998)]{ferguson1998computer}
Ferguson, J. Computer programs for assessment of long-wavelength radio communications, version 2.0: User's guide and source files. Tech. Rep., DTIC Document (1998). 

\bibitem[{\textit{Foullon et~al.}}(2005)]{foullon2005}
Foullon, C., Verwichte, E., Nakariakov, V. M., Fletcher, L., X-ray quasi-periodic pulsations in solar flares as magnetohydrodynamic oscillations. \textit{Astronomy and Astrophysics} 440, L59-L62 (2005).



\bibitem[{\textit{Gledhill}}(1986)]{gledhill}
Gledhill, J., The effective recombination coefficient of electrons in the ionosphere between 50 and 150 km. \textit{Radio Science} 21, 399-408 (1986)



\bibitem[{\textit{Grubor et~al.}}(2005)]{grubor05}
Grubor, D., {\v{S}}uli{\'c} D., {\v{Z}}igman, V., Influence of Solar X-ray Flares on the Earth-Ionosphere Waveguide. \textit{Serbian Astronomical Journal} 171, 29-35 (2005). 


\bibitem[{\textit{Harrison}}(1987)]{harrison1987solar}
Harrison, R. Solar soft X-ray pulsations. \textit{Astronomy and Astrophysics} 182, 337-347 (1987). 



\bibitem[{\textit{Inan et~al.}}(1999)]{inan1999}
Inan, U. \textit{et al.} Ionization of the Lower Ionosphere by $\gamma$-rays from Magnetar: Detection of a Low Energy (3-10 kev) Component. \textit{Geophysical Research Letters} 26, 3357-3360 (1999). 



\bibitem[{\textit{Inan et~al.}}(2010)]{inan2010_lightening}
Inan, U. S., Cummer, S. A., Marshall, R. A. A survey of ELF and VLF research on lightning-ionosphere interactions and causative discharges. \textit{Journal of Geophysical Research: Space Physics} 115, (2010). 

\bibitem[{\textit{Kar et~al.}}(1986)]{kar}
Kar, J., Mahajan, K., Srilakshmi, M., Kohli, R. Possible effects of solar flares on the ionosphere of Venus from pioneer Venus orbiter measurements. \textit{Journal of Geophysical Research: Space Physics} 91, 8986-8992 (1986).

\bibitem[{\textit{Kaufmann et~al.}}(1989)]{kaufmann1989effects}
Kaufmann, P. \textit{et al.} Effects of the large June 1975 meteoroid storm on Earth's ionosphere. \textit{Science} 246, 787-791 (1989).



\bibitem[{\textit{Kumar et~al.}}(2015)]{kumar2015response}
Kumar, S. et al. Response of the low-latitude D-region ionosphere to extreme space weather event of 14–16 December 2006. \textit{Journal of Geophysical Research: Space Physics} 120, 788-799 (2015).

\bibitem[{\textit{Kumar et~al.}}(2016)]{kumar2016changes}
Kumar, S., Kumar, A., Maurya, A. K., Singh, R. Changes in the D-region associated with three recent solar eclipses in the south pacific region. \textit{Journal of Geophysical Research: Space Physics} 121, 5930-5943 (2016). 



\bibitem[{\textit{Lemen et~al.}}(2011)]{lemen}
Lemen, J. R. et al. The Atmospheric Imaging Assembly (AIA) on the Solar Dynamics Observatory (SDO). \textit{Solar Physics}, 275(1), 17-40 (2011).

\bibitem[{\textit{Li \& Gan}}(2008)]{li}
Li, Y. P., \& Gan, W. Q. (2008). Observational studies of the X-ray quasi-periodic oscillations of a solar flare. \textit{Solar Physics}, 247(1), 77-85.

\bibitem[{\textit{Mendillo et~al.}}(2006)]{mendillo06}
Mendillo, M., Withers, P., Hinson, D., Rishbeth, H., Reinisch, B., Effects of solar flares on the ionosphere of Mars. \textit{Science} 311, 1135-1138 (2006). 


\bibitem[{\textit{Mitra}}(1974)]{mitra74}
Mitra, A. Ionospheric Effects of Solar Flares.
D. Reidel Publishing Company, Dordrecht, Holland (1974)


\bibitem[{\textit{Mitra-Kraev et~al.}}(2005)]{mitra2005first}
Mitra-Kraev, U., Harra, L., Williams, D., Kraev, E. The first observed stellar X-ray flare oscillation: constraints on the flare loop length and the magnetic field. \textit{Astronomy and Astrophysics} 436, 1041-1047 (2005). 



\bibitem[{\textit{Mumford et~al.}}(2015)]{mumford2015sunpy}
Mumford, S. J. \textit{et al.} Sunpy - Python for solar physics. \textit{Computational Science \& Discovery} 8, 014009 (2015). 

\bibitem[{\textit{Nakariakov and Melnikov}}(2009)]{nak2009}
Nakariakov, V. M., Melnikov, V. Quasi-periodic pulsations in solar flares. \textit{Space Science Reviews} 149, 119-151 (2009) 

\bibitem[{\textit{Nakariakov et~al.}}(2016)]{nakariakov2016magnetohydrodynamic}
Nakariakov, V. M. et al. Magnetohydrodynamic oscillations in the solar corona and Earth's magnetosphere: Towards consolidated understanding. \textit{Space Science Reviews} 200, 75-203 (2016). 




\bibitem[{\textit{Nina et~al.}}(2012)]{nina2012altitude}
Nina, A., {\v{C}}ade{\v{z}}, V., Sre{\'c}kovi{\'c}, V., {\v{S}}uli{\'c}, D., Altitude distribution of electron concentration in ionospheric D-region in presence of time-varying solar radiation flux. \textit{Nuclear Instruments and Methods in Physics Research Section B: Beam Interactions with Materials and Atoms} 279, 110-113 (2012). 


\bibitem[{\textit{Nina et~al.}}(2013)]{nina2013detection}
Nina, A., {\v{C}}ade{\v{z}}, V. Detection of acoustic-gravity waves in lower ionosphere by VLF radio waves. \textit{Geophysical Research Letters} 40, 4803-4807 (2013). 





\bibitem[{\textit{Parrot and Mogilevsky}}(1989)]{earthquake}
Parrot, M., Mogilevsky, M. VLF emissions associated with earthquakes and observed in the ionosphere and the magnetosphere. \textit{Physics of the Earth and Planetary Interiors} 57, 86-99 (1989). 

\bibitem[{\textit{Pugh et~al.}}(2016)]{pugh2016statistical}
Pugh, C., Armstrong, D. J., Nakariakov, V. M., Broomhall, A. M. Statistical properties of quasi-periodic pulsations in white-light flares observed with Kepler. \textit{Monthly Notices of the Royal Astronomical Society} 459, 3659-3676 (2016).

\bibitem[{\textit{Ratcliffe}}(1972)]{ratcliffe}
Ratcliffe, J. A. Introduction to the Ionosphere and Magnetosphere, \textit{Cambridge University Press, Cambridge} 1972). 

\bibitem[{\textit{Selvakumaran et~al.}}(2015)]{selv}
Selvakumaran, R., Maurya, A. K., Gokani, S. A., Veenadhari, B., Kumar, S., Venkatesham, K., \& Singh, R. (2015). Solar flares induced D-region ionospheric and geomagnetic perturbations. \textit{Journal of Atmospheric and Solar-Terrestrial Physics}, 123, 102-112.

\bibitem[{\textit{Scherrer et~al.}}(2008)]{scherrer2008distributing}
Scherrer, D. et al., Distributing space weather monitoring instruments and educational materials worldwide for IHY 2007: The AWESOME and SID project. \textit{Advances in Space Research} 42, 1777-1785 (2008). 



\bibitem[{\textit{Shao et~al.}}(2013)]{nat_ion}
Shao, X.-M., Lay, E. H. \& Jacobson, A. R. Reduction of electron density in the night-time lower ionosphere in response to a thunderstorm. \textit{Nature Geoscience} 6 29-33 (2013)



\bibitem[{\textit{\v{S}vestka et~al. }}(1982)]{vsvestka1982unusual}
{\v{S}}vestka, Z. et al. Unusual coronal activity following the flare of 6 November 1980. \textit{Solar Physics} 80, 143-159 (1982). 



\bibitem[{\textit{\v{S}vestka}}(1994)]{vsvestka1994slow}
{\v{S}}vestka, Z. Slow-mode oscillations of large-scale coronal loops. \textit{Solar Physics} 152, 505-508 (1994). 




\bibitem[{\textit{Tan et~al.}}(2016)]{tan2016very}
Tan, B., Yu, Z., Huang, J., Tan, C., Zhang, Y., Very long-period pulsations before the onset of solar flares. \textit{The Astrophysical Journal} 833, 206  (2016).



\bibitem[{\textit{Thomson and Clilverd}}(2001)]{thomson2001solar}
Thomson, N. R., Clilverd, M. A. Solar flare induced ionospheric D-region enhancements from VLF amplitude observations. \textit{Journal of Atmospheric and Solar-Terrestrial Physics} 63, 1729-1737 (2001). 

\bibitem[{\textit{Thomson et~al.}}(2005)]{thomson2005large}
Thomson, N. R., Rodger, C. J., Clilverd, M. A. Large solar flares and their ionospheric D region enhancements. \textit{Journal of Geophysical Research: Space Physics} 110, (2005).



\bibitem[{\textit{Tsurutani et~al.}}(2009)]{tsurutani2009brief}
Tsurutani, B.T., Verkhoglyadova, O.P., Mannucci, A.J., Lakhina, G.S., Li, G. and Zank, G.P., 2009. A brief review of `solar flare effects' on the ionosphere. \textit{Radio Science}, 44(1).


\bibitem[{\textit{Van Doorsselaere et~al.}}(2016)]{vand}
Van Doorsselaere, T., Kupriyanova, E. G., \& Yuan, D. (2016). Quasi-periodic pulsations in solar and stellar flares: an overview of recent results (Invited review). \textit{Solar Physics}, 291(11), 3143-3164.


\bibitem[{\textit{Wait and Spies}}(1964)]{wait1964}
Wait, J. R. \& Spies, K. P. Characteristics of the Earth-ionosphere waveguide for VLF radio waves. 300 (US Dept. of Commerce, National Bureau of Standards, 1964). 

\bibitem[{\textit{Wang}}(2011)]{wang}
Wang, T. (2011). Standing slow-mode waves in hot coronal loops: observations, modeling, and coronal seismology. \textit{Space science reviews}, 158(2), 397-419.


\bibitem[{\textit{Wenzel et~al.}}(2016)]{gifds} 
Wenzel, D., Jakowski, N., Berdermann, J., Mayer, C., Valladares, C., \& Heber, B. (2016). Global ionospheric flare detection system (GIFDS). \textit{Journal of Atmospheric and Solar-Terrestrial Physics}, 138, 233-242.


\bibitem[{\textit{Whitten and Poppoff}}(1965)]{whitten1965physics}
Whitten, R. C. \& Poppoff, I. G., Physics of the Lower Ionosphere Prentice Hall, New Jersey 1965. 

\bibitem[{\textit{Witasse et~al.}}(2008)]{witasse} Witasse, O., Cravens, T., Mendillo, M., Moses, J., Kliore, A., Nagy, A.F. and Breus, T., 2008. Solar system ionospheres. In Comparative Aeronomy (pp. 235-265). Springer New York. Solar System Ionospheres.
\textit{Space Science Reviews} 139 235-265 (2008)



\bibitem[{\textit{Zigman et~al.}}(2007)]{zigman2007}
{\v{Z}}igman, V., Grubor, D., {\v{S}}uli{\'c}, D. D-region electron density evaluated from VLF amplitude time delay during X-ray solar flares. \textit{Journal of atmospheric and solar-terrestrial physics} 69, 775-792 (2007). 











 \end{thebibliography}
%
%
%
%





\listofchanges

\end{document}